\documentclass[namedreferences]{SolarPhysics}
\usepackage[optionalrh]{spr-sola-addons} % For Solar Physics

\usepackage[pdfborder={0 0 0},urlcolor=blue,breaklinks]{hyperref}
\usepackage{epsfig}          % For eps figures, old commands
\usepackage{graphicx}        % For eps figures, newer & more powerfull
\usepackage{courier}         % Change the \texttt command to courier style
\usepackage{amssymb}        % useful mathematical symbols
\usepackage{color}           % For color text: \color command

\newcommand{\be}{\begin{equation}}
\newcommand{\ee}{\end{equation}}
\newcommand{\beq}{\begin{eqnarray}}
\newcommand{\eeq}{\end{eqnarray}}

\usepackage{stdclsdv}
\usepackage{enumerate}

\newcommand{\jgr}{    {\it J. Geophys. Res.}}

\begin{document}
\begin{article}
\begin{opening}

\title{The Electron Temperature and Anisotropy in the Solar Wind. 
I. Comparison of the Core and Halo Populations}
\author{V. \surname{Pierrard}$^{1,2}$ $\sep$M. \surname{Lazar}$^{3,4}$ $\sep$
S. \surname{Poedts}$^{4}$ $\sep$\v{S}.
\surname{\v{S}tver\'{a}k}$^{5,6}$ $\sep$M. \surname{Maksimovic}$^{7}$ $\sep$P.M. 
\surname{Tr\'avn\'i\v{c}ek}$^{5,6,8}$ 
}

\runningauthor{V. Pierrard \emph{et al.}} \runningtitle{Temperature
Anisotropy of the Solar Wind Electrons. I.}

\institute{$^1$Royal Belgian Institute for Space Aeronomy, Space Physics and Solar-Terrestiral Center of Excellence, 3 av. Circulaire, B-1180 Brussels, Belgium \\
$^2$Universit\' e Catholique de Louvain (UCL), Center for Space Radiations (CSR) and Georges Lema\^ itre Centre for Earth and Climate Research (TECLIM), Earth and Life Institute (ELI), Place Louis Pasteur 3 bte L4.03.08, B-1348 Louvain-La-Neuve, Belgium\\
$^3$Institut f\"ur Theoretische Physik, Lehrstuhl IV: Weltraum- und Astrophysik, Ruhr-Universit\"at Bochum, Germany\\
$^4$Center for Mathematical Plasma Astrophysics, K.U. Leuven, Celestijnenlaan 200B, 3001 Leuven, Belgium\\
$^5$Institute of Atmospheric Physics, Czech Academy of Sciences, Prague, Czech Republic\\
$^6$Astronomical Institute, Czech Academy of Sciences, Ondrejov, Czech Republic\\
$^7$LESIA and CNRS, Observatoire de Paris-Meudon, Meudon, France\\
$^8$Space Science Laboratory, University of California
\email{viviane.pierrard@aeronomie.be}}
\date{Received ; accepted }

\begin{abstract}
Estimating the temperature of the solar wind particles and their anisotropies is particularly
important for understanding the origin of these deviations from thermal equilibrium as well as 
their effects. In the absence of energetic events the velocity distribution of electrons reveal 
a dual structure with a thermal (Maxwellian) core and a suprathermal (Kappa) halo.
This paper presents a detailed observational analysis of these two components, providing 
estimations of their temperatures and temperature anisotropies and decoding any potential 
interdependence that their properties may indicate. The data set used in this study 
includes more than 120$\,$000 the events detected by three missions in the ecliptic within an extended
range of heliocentric distances from 0.3 to over 4~AU. The anti-correlation found for the core 
and halo temperatures is consistent with the radial evolution of the Kappa model, clarifying an
apparent contradiction in previous observational analysis and providing valuable clues about the
temperature of the Kappa-distributed populations. However, these two components manifest a clear 
tendency to deviate from isotropy in the same direction, that seems to confirm the existence of 
mechanisms with similar effects on both components, e.g., the solar wind expansion, or the particle 
heating by the fluctuations. On the other hand, the existence of plasma states with anti-correlated 
anisotropies of the core and halo populations suggests a dynamic interplay of these components, 
mediated, most probably, by the anisotropy-driven instabilities.

\end{abstract}


\end{opening}

%%%%%%%%%%%%%%%%%%%%%%%%%%%%%%%%%%%%%%%%%%%%%%%%%%%%%%%%%%%%%%%%%%%%%%%
\section{Introduction}\label{Introduction}
%%%%%%%%%%%%%%%%%%%%%%%%%%%%%%%%%%%%%%%%%%%%%%%%%%%%%%%%%%%%%%%%%%%%%%%

Ejected by the Sun in space the solar wind is an expanding plasma
comprising electrons, protons and minor ions (e.g., He, O, N). At
1~AU where it is frequently observed, this plasma is very hot
($T_{\rm particle} = 10^5 - 10^6$ K) and very rare at the same time
($n = 1-10$ cm$^{-3}$), such that it requires a kinetic (comprehensive) description
based on the in-situ measurements of the particle velocity
distributions of the components. Measured as particle fluxes, the
velocity distributions reveal two major components, a
dominant low-energy core ($> 90 \% n_{\rm total}$) and a
suprathermal halo that enhances the high-energy tails of the
velocity distributions and is well described by the power-law
Kappa-like distribution functions \cite{Pier01, ma05, pi10}. Although
the halo is present in all species of plasma particles, i.e., 
electrons, protons or heavier ions \cite{va68, ch89, co96, st08} as a 
distinct component markedly different from the core, the origin of these
suprathermal populations is still uncertain. Multiple plausible
scenarios have been proposed, including a coronal origin \cite{sc92, Pier99,
vi00} of the suprathermals, as well as their generation by
heating the resonant plasma particles by the fluctuations \cite{ma98} 
transported by the super-Alfv\'enic solar wind, or the isotropisation of the
field-aligned streams (also known as strahls) by the selfgenerated
instabilities \cite{ma05, ga07, pa07, Pier11, vo12, pa13}. Not only the
origin of these two components is still controversial, but also
their interplay and subsequent implication in the solar wind
dynamics remains unclear \cite{li98, la12b}.

In addition to the number density criteria mentioned above, the temperature 
and its anisotropy $T_\parallel \ne T_\perp$, in general with respect to the
uniform magnetic field, can also be invoked as macroscopic parameters to separate 
and describe the core and halo components. The magnitude of the temperature 
anisotropy may provide important clues on the macro- or microscopic mechanisms 
at work in the solar wind. For instance, the observations do not confirm an excessive 
increase of the temperature of plasma particles in direction parallel to the 
interplanetary magnetic field as an effect of adiabatic expansion of the solar wind.
The observed limits of the temperature anisotropy may be an effect of the microscopic processes, 
like particle-particle or wave-particle interactions, which prevent large deviations 
from isotropy. The observational analysis proposed in the present paper attempts therefore to unveil 
detailed features of the temperature anisotropy of the electron core and halo components, and 
create premises for a realistic interpretation of these physical processes. % as one of the main sources of free energy 
%and kinetic instabilities, which may explain the enhanced fluctuations frequently observed
%in the solar wind, and, implcitly, the constraints imposed by these instabilities to the temperature anisotropy. 

Although the first reliable information on the temperature anisotropy of the solar 
wind electrons may be dated back in '70s \cite{se72, sc73, fe75}, the number of 
subsequent reports with systematic studies is relatively limited \cite{ph89, ph90, pi90, st08}. 
In the early studies the electron temperature and the temperature anisotropy were 
determined by fitting the velocity distribution measured in-situ with standard 
bi-Maxwellian models. The core temperatures $T_{c, \parallel}$ and $T_{c, \perp}$, 
determined by bi-Maxwellian fits in directions parallel and perpendicular to the 
magnetic field, respectively, decrease faster with distance from the Sun than the 
components of the total temperature $T_{\parallel}$ and $T_{\perp}$ as determined 
from the entire electron distributions \cite{pi90}. At higher energies, the halo 
electrons were examined by fitting the suprathermal tails of the distribution 
to a second bi-Maxwellian with a lower density but higher temperatures \cite{fe75, 
ph89,ma00}, or to a bi-Kappa also known as a bi-Lorentzian distribution function 
\cite{ma05,st08}.

Correlations with plasma stream structures indicate that the electron temperatures 
are generally higher in the slow solar wind than in high-speed stream \cite{pi90}. An excess 
of parallel temperature $T_{\parallel} > T_{\perp}$ has been observed to dominate 
the observations and it is significantly larger in high-speed streams than in the 
slow solar wind, while an excess of perpendicular temperature $T_\perp > T_\parallel$ 
is more specific to low speed and high density conditions. The core anisotropy 
also varies systematically as a function of electron density, being diminished 
with increasing the density \cite{ph89,ph90}. It was also established a direct interdependence between
the self-collisional frequency and the anisotropy of electrons \cite{sa03}.
%where $T_{e,\parallel} / T_{e,\perp}$ take typical values in the interval 1.5-1.7.
The existing information about the anisotropy of suprathermal electrons
is less detailed compared to the core. Limitations in this case may also result from 
describing the halo with bi-Maxwellian models which cannot reproduce accurately the 
suprathermal tails, and from the fact that all these analyses were restrained to a 
reduced number of events \cite{fe75,pi87,pi90,ph89}. However, it seems to be evident 
that the halo component is more anisotropic than the core, a result
in agreement with the assumption that suprathermal electrons have Coulomb scattering 
times much longer than the core electrons, and comparable to their transit times to 
1~AU. Similar to the core, the electron halo exhibits in general an excess of
parallel temperature $T_{h,\parallel} > T_{h,\perp}$ which is enhanced with the increase
of the solar wind flowing speed, most probably because of the field aligned strahl
which is frequently included in the halo component.

Many of these preliminary results are re-confirmed by the more advanced studies which provide a 
detailed parametrization of the solar wind electrons by fitting the core component to 
a bi-Maxwellian, the halo to a bi-Kappa and the strahl to a drifting bi-Kappa 
\cite{ma05,st08}. These studies consider a large number of events combining solar wind 
data from different spacecraft missions, e.g., Helios I and II, Ulysses, Wind, and 
Cluster II, and within a wide range of heliographic coordinates. More pronounced in the fast winds 
the field-aligned strahl is diminished with heliocentric distance, especially beyond 1 AU, 
while the halo component is enhanced, suggesting the existence of a scattering mechanism that 
isotropize the strahl population \cite{ma05}. An enhancement of the suprathermal populations
quantified by a decrease of the power-index kappa (noted $\kappa$ in the next) with radial distance
from the Sun is also observed, but this result seems to be in contradiction with the same trend 
of decreasing found for the Kappa (halo) temperature.
On the other hand, a dual model combining a bi-Maxwellian core and a bi-Kappa halo is 
sufficient to parameterize the anisotropic electrons not only in the slow wind, but 
for any fast or slow wind conditions beyond 1 AU. \inlinecite{st08} have invoked
a dual halo-core approach to build statistical histograms of the temperature anisotropy of 
these two components for the slow and fast wind conditions in the ecliptic.
The limits of the temperature anisotropy for the core and halo populations 
are provided for the full range of electron plasma beta parameter, enabling 
comparative studies with the constraints predicted by the resulting instabilities, 
Coulomb collisions, or adiabatic expansion of the solar wind. However, \inlinecite{st08}
have limited to study in detail only the core anisotropy, showing that greater temperature
anisotropies can develop at larger distances from the Sun, and collisions still may 
have an effect to maintain low levels of temperature anisotropy complementing the 
kinetic instabilities which constrain large deviations from isotropy  of the core population.

In the present paper we report the results of a supplementary extended investigation 
on the same set of data sampled in \inlinecite{st08}. Our main aim is to complete the 
puzzle by exploring the physical features of the suprathermal halo electrons, 
which here are presented by contrast to the properties of the core population. 
A comprehensive picture of these populations and their properties is expected to provide 
answers to the questions that are still open as well as premises for understanding
the origin of these populations and their implication in key processes like particle
heating or particle and energy transport in the solar wind.

%%%%%%%%%%%%%%%%%%%%%%%%%%%%%%%%%%%%%%%%%%%%%%%%%%%%%%%%%%%%%%%%%%%%%%%%%%%%%%%%
\section{Fitting models for the velocity distributions}
%%%%%%%%%%%%%%%%%%%%%%%%%%%%%%%%%%%%%%%%%%%%%%%%%%%%%%%%%%%%%%%%%%%%%%%%%%%%%%%%

The data set we propose to analyze in these series of papers are the main plasma
parameters which describe the electron populations, namely,
number density ($n$), temperature ($T$), temperature anisotropy
($A= T_\perp/T_\parallel$), plasma beta ($\beta = 8 \pi n k_B T / B_0^2$),
and the power-index $\kappa$ to quantify the presence of suprathermal halo
population. These parameters were estimated by fitting the
observed velocity distributions with a dual core-halo
(subscripts $c$ and $h$, respectively) analytical model \cite{ma05}
\begin{equation} f(v_{\parallel}, v_{\perp}) = f_{c}
(v_{\parallel}, v_{\perp}) + f_{h}(v_{\parallel}. v_{\perp})
\label{e1} \end{equation}
Such a model may describe these two principal components even in the presence of a
strahl (beaming) component, which is present especially in the fast winds ($V_{\rm SW} > 400$ km/s).
With an anti-Sunward orientation, the field-aligned strahl overlaps asymmetrically in the velocity distribution
\cite{pi87,og00,st08, vi10} and can therefore be subtracted. In the
present study we focus our attention exclusively on the core and
halo populations of electrons, which are always present in the observations. Moreover,
the halo is significantly enhanced with heliographic distance while the strahl is diminished
in the same measure \cite{ma05}.

Both the core and halo components can be anisotropic with respect to the magnetic field direction,
but their distributions are assumed gyrotropic and well approximated in polar coordinates
$(v_x, v_y, v_z)=(v_{\perp} \cos \phi, v_{\perp} \sin \phi, v_{\parallel})$ by two-axis
distribution functions. Thus, the core is described by a bi-Maxwellian
\begin{equation} f_c (v_{\parallel}, v_{\perp}) = \left(m \over 2 \pi k_B\right)^{3/2} {n_c \over
T_{c, \parallel}^{1/2} T_{c,\perp}} \, \exp \left[- {m\over 2 k_B} \left({v_{\parallel}^2\over
T_{c,\parallel} }+ {v_{\perp}^2\over T_{c,\perp} }\right) \right],
\label{e2} \end{equation}
while the halo is best fitted by a bi-Kappa
\beq f_h (v_{\parallel}, v_{\perp})  & = & \left(m \over (2\kappa-3)\pi k_B\right)^{3/2}
{n_h \over T_{h,\parallel}^{1/2} T_{h,\perp}} \, {\Gamma[\kappa + 1] \over
\Gamma[\kappa - 1/2]} \nonumber \\ 
&& \times \left[1 + {m\over (2\kappa-3) k_B} \left({v_{\parallel}^2\over
T_{h,\parallel}} + {v_{\perp}^2\over T_{h,\perp}^2}\right)\right]^{-\kappa-1}. \label{e3} \eeq
Fitting parameters are the number densities of these two
components, $n_c$ and $n_h$, and the temperature components as
moments of second order
\begin{equation} T_{c,\parallel} = {m \over n_c k_B} \int d{\bf v}
v_{\parallel}^2 f_c, \;\;\; T_{c,\perp} = {m \over 2 n_c k_B} \int d{\bf v}
v_{\perp}^2 f_c,
\label{e4} \end{equation}
\begin{equation} T_{h,\parallel} = {m \over n_h k_B} \int d{\bf v}
v_\parallel^2 f_h, \;\;\; T_{h,\perp} = {m \over 2 n_h k_B} \int d{\bf v}
v_\perp^2 f_h\label{e5} \end{equation}
In the limit of a very large power-index $\kappa \to \infty$ the bi-Kappa
reduces to a bi-Maxwellian, leading to a standard description for both the core and halo components, see the
two-Maxwellian model in \inlinecite{fe75} or \inlinecite{ma97}.
These details on the velocity distribution functions will help understanding the
results of our study in the next, but first we reintroduce the observational data
and discuss their implications.

%%%%%%%%%%%%%%%%%%%%%%%%%%%%%%%%%%%%%%%%%%%%%%%%%%%%%%%%%%%%%%%%%%%%%
\section{Observational data}
%%%%%%%%%%%%%%%%%%%%%%%%%%%%%%%%%%%%%%%%%%%%%%%%%%%%%%%%%%%%%%%%%%%%%

\inlinecite{st08} have fitted the electron velocity distributions
for, roughly, 124,000 events at low latitudes (in the ecliptic).
These events were collected by three spacecraft as follows, see also
Table I in \inlinecite{st08}: (1) $\sim 100,000$ samples from Helios 1,
in the time interval 1975 - 1978, and small heliocentric distances
0.3 - 1.0 AU; (2) $\sim 10,000$ samples from Cluster II, in a two-years
interval 2002 - 2003, at 1 AU; and (3) $\sim 14,000$ samples from Ulysses
in another two-years interval 1990 - 1991, and large distances 1.2 - 3.95 AU.
Detailed descriptions for the electron analyzers used by these three
missions, and for the methods of correction and reconstruction
of the full 3D electron velocity distribution functions are provided
in \inlinecite{st08}. In the presence of strahl the observed distributions
are asymmetric but only data points measured in the sunward direction were fitted
to the model (\ref{e1}), assuming that both the core and halo
populations have the same properties in the antisunward direction.
Data below the one count level were removed. The other plasma
parameters invoked in our study, e.g., solar wind bulk speed,
heliocentric distance, and the ambient magnetic field, are taken
from measurements made by the plasma instruments on board these
three spacecraft.

This data set has been used for studying the observed limits of the
electron temperature anisotropy for the core and halo populations
\cite{st08}, in an endeavor to explain their physical constraints.
These limits were compared with the thresholds of the temperature
anisotropy driven instabilities: For the Maxwellian core in the slow
wind the limits of the temperature anisotropy appear to be well
shaped by the instability thresholds predicted for the anisotropic
bi-Maxwellian distributions, see top panel in Fig.~5 from
\inlinecite{st08}. However, the same instability thresholds derived
for bi-Maxwellian distributed plasmas are not relevant for the halo
component, see Fig.~6 in \inlinecite{st08}, because the main
quantities describing the halo populations (\emph{i.e.}, particle
number density, temperatures, anisotropy) are determined by fitting
their distributions with anisotropic bi-Kappa models, which
reproduce a bi-Maxwellian only in the limit of a very large
power-index $\kappa \to \infty$ (\emph{i.e.}, in the absence of
suprathermal populations).

%The electron velocity distributions were measured for, roughly,
%250,000 events at low latitudes (in the ecliptic), see Table I in
%\inlinecite{st09} [an extended work of the previous analysis of only
%120,000 events by \inlinecite{st08}], and these events were
%collected by three spacecraft as follows: (1) $\sim 200,000$ samples
%from Helios I, in the time interval 1975 - 1978, and small
%heliocentric distances 0.3 - 1.0 AU; (2) $\sim 25,000$ samples from
%Cluster II, in a two-years interval 2002 - 2003, at 1 AU; and (3)
%$\sim 15,000$ samples from Ulysses in another two-years interval
%1990 - 1991, and large distances 1.2 - 3.95 AU. Detailed
%descriptions are provided in \inlinecite{st08} and \inlinecite{st09}
%on the electron analyzers used by these three missions as well as
%the methods of correction and reconstruction of the full 3D electron
%velocity distribution functions.

\begin{figure}[h]
%\centerline{
    \includegraphics[width=60mm]{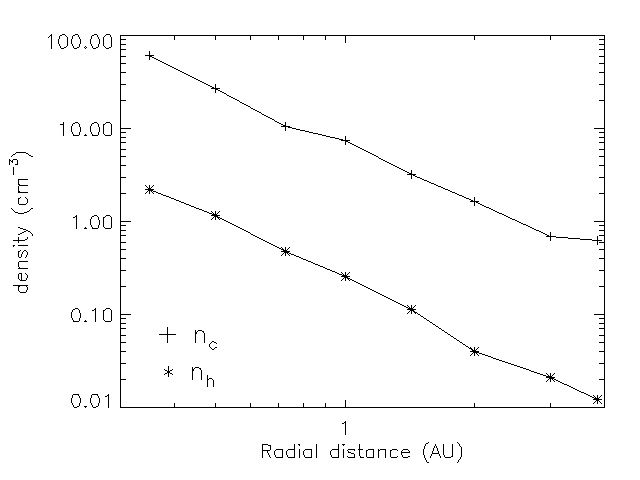}
    \includegraphics[width=60mm]{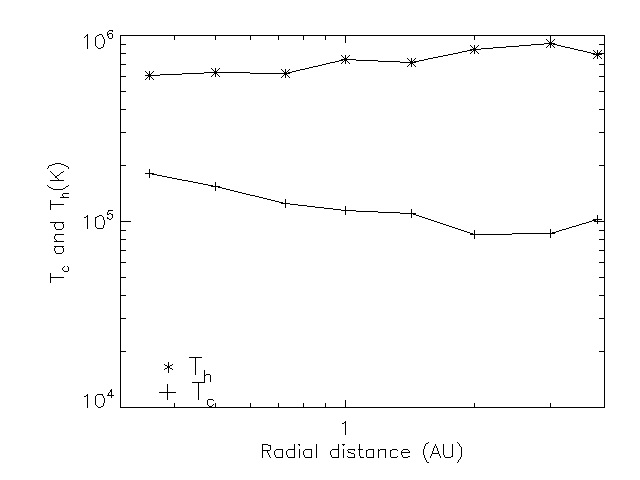}\\
    \includegraphics[width=60mm]{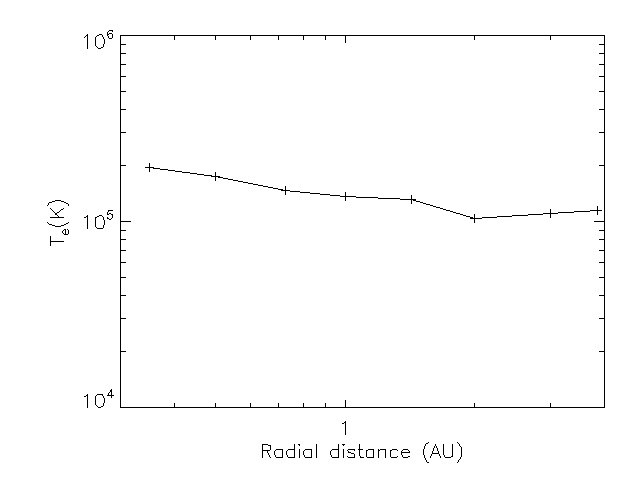}
    \includegraphics[width=60mm]{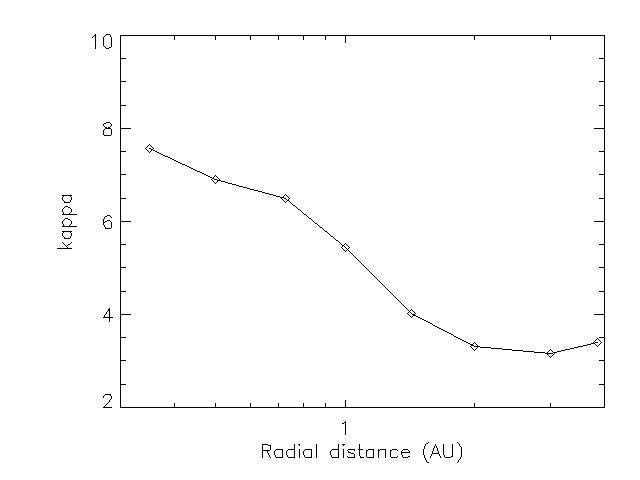}
\caption{Radial profile of the electron parameters in the core
(cross) and halo (star): (a) Electron number densities, $n_c$ and
$n_h$, (b) total temperature $T_e = (n_c T_c + n_h T_h)/(n_c + n_h)$, 
(c) temperatures, $T_c = (2T_{c,\perp}+T_{c,\parallel})/3$
and $T_h = (2T_{h,\perp}+T_{h,\parallel})/3$,  and (d) the power-index
$\kappa$ for the halo component.} \label{f1}
\end{figure}

The same analytical model in Eq. (\ref{e1}) has been used by
\inlinecite{ma05} for a different set of data combining events from
the ecliptic detected by Helios 1 and 2 (1976) and Wind (1995), and
events reported by Ulysses (1995) from time intervals corresponding
to the first south and north polar passes. They present a radial
evolution of the electron plasma components from 0.3 to 1.5 AU, building radial
profiles for their main parameters, e.g., temperatures for the core
($T_c$) and halo ($T_h$) populations, and number densities for any
of the core ($n_c$), halo ($n_h$) and strahl ($n_s$) components (the
strahl density is obtained by subtracting the core and halo
electrons from total distribution). It was thus found that the core
fractional density remains roughly constant with heliocentric
distance, while the halo and strahl fractional densities vary in an
opposite way: the relative number of halo electrons is increasing on
an apparent expense of the strahl electrons which decrease with
radial distance. Moreover, the radial profile of the power-index
$\kappa$ indicates a decrease from about $\kappa = 6 - 7$ at $0.3 -
0.4$~AU to $\kappa \gtrsim 3$ at $1.35 - 1.5$~AU \cite{st09}. These are the first 
important evidences confirming the scenario that
the heliospheric electron halo population consists partly of
electrons that have been scattered out of the strahl, also
suggesting that kinetic instabilities induced by the strahl (e.g.,
electrostatic or electromagnetic beam-plasma instabilities) may be a
source of pitch-angle scattering dominating the magnetic focusing of
the strahl and enhancing the halo population \cite{pa07,ga07,vo12}.

\begin{table}[h]
\caption{Mean values for the plasma parameters determined in the solar wind,
and used to build their heliocentric radial profiles in Figs. \ref{f1} - \ref{f4}. 
The intervals (bins) adopted in the calculation are the following (in AU): 0.3-0.4; 
0.475-0.53; 0.7-0.75; 0.9-1.1; 1.35-1.5; 1.9-2.1; 2.9-3.1; 3.75-3.95.}
\label{tab1}
\begin{tabular}{c c c c c c c c c}\hline
Distance (AU) &  0.35 & 0.5 & 0.725 & 1.0 & 1.425 & 2.0 & 3.0 & 3.85 \\
\hline $n_c (cm^{-3})$ & 61.14 & 26.60 & 10.51 & 7.34 & 3.22 & 1.64 & 0.69 & 0.42\\
\hline $n_h$  & 2.19 & 1.16 & 0.48 & 0.26 & 0.11 & 0.04 & 0.02 & 0.01\\
\hline $T$(10$^5$ K) & 1.962 & 1.749 & 1.463 & 1.360 & 1.307 & 1.033 & 1.105 & 1.170 \\
\hline $\kappa$ & 7.57 & 6.89 & 6.49 & 5.43 & 4.02 & 3.31 & 3.16 & 3.40\\
\hline $u$ (km/s)& 440 & 467 & 489 & 464 & 384 & 380 & 399 & 430\\
\hline $T_c$ (10$^5$ K) & 1.814 & 1.550 & 1.246 & 1.145 & 1.105 & 0.852 & 0.863 & 1.020 \\
\hline $T_{c,\parallel}$(10$^5$ K) & 1.944 & 1.702 & 1.348 & 1.259 & 1.186 & 0.861 & 0.909 & 1.083 \\
\hline $T_{c,\perp}$ (10$^5$ K) & 1.749 & 1.474 & 1.194 & 1.088 & 1.064 & 0.848 & 0.839 & 0.956 \\
\hline $A_c (\kappa = 3)$ &  - & 0.86 & 0.74 & 0.89 & 0.87 & 0.99 & 0.87 & 0.87 \\
\hline $A_c (\kappa = 5)$ & 0.80 & 0.76 & 0.83 & 0.88 & 1.01 & 0.99 & 1.02 & 1.14 \\
\hline $A_c (\kappa = 7)$ & 0.87 & 0.87 & 0.91 & 0.90 & - & - & - & -\\
\hline $T_h$ (10$^5$ K) & 6.092 & 6.313 & 6.243 & 7.467 & 7.148 & 8.388 & 9.070 & 7.857 \\
\hline $T_{h,\parallel}$(10$^5$ K) & 6.375 & 6.581 & 6.421 & 7.618 & 7.927 & 8.391 & 8.830 & 8.113 \\
\hline $T_{h,\perp}$(10$^5$ K) & 5.951 & 6.179 & 6.154 & 7.391 & 6.759 & 8.386 & 9.190 & 7.960 \\
\hline $A_h (\kappa = 3)$ & - & 1.10 & 0.84 & 0.96 & 1.02 & 1.02 & 0.97 & 0.95 \\
\hline $A_h (\kappa = 5)$ & 0.78 & 0.85 & 0.92 & 0.99 & 0.92 & 0.99 & 1.03 & 1.11 \\
\hline $A_h (\kappa = 7)$ & 0.93 & 0.97 & 1.01 & 1.00 & 0.94 & - & - & - \\
\hline
\end{tabular}
\end{table}

%%%%%%%%%%%%%%%%%%%%%%%%%%%%%%%%%%%%%%%%%%%%%%%%%%%%%%%%%%%%%%%%%%%%%%%%%
\section{Comparative analysis: halo vs. core}
%%%%%%%%%%%%%%%%%%%%%%%%%%%%%%%%%%%%%%%%%%%%%%%%%%%%%%%%%%%%%%%%%%%%%%%%%

Here we present the results of our comparative analysis on the main
properties of the electron core and halo populations distinctively
provided by the moments of the velocity distributions measured at
low latitudes in the solar wind. Mean values of the plasma parameters
used in our analysis are listed in Tables~\ref{tab1} and \ref{tab2}. For the 
radial evolution (Table~\ref{tab1}) the average values are calculated
for 8 bins scalled in AU units: 0.3-0.4; 0.475-0.53; 0.7-0.75; 0.9-1.1; 
1.35-1.5; 1.9-2.1; 2.9-3.1; 3.75-3.95, where the first five radial bins are
chosen as in \inlinecite{ma05}. For a $\kappa$-dependency (Table~\ref{tab2}) the average values
are determined for 13 bins: $<$2.75; 2.75-3.25; 3.25-3.75; 3.75-4.25; 4.25-4.75; 
4.75-5.25; 5.25-5.75; 5.75-6.25; 6.25-6.75; 6.75-7.25; 7.25-7.75; 7.75-8.25; $>$8.25.

The events invoked by \inlinecite{ma05} were restricted 
to heliocentric distances between 0.3 and 1.5~AU, and to the fast wind 
steady states, selecting only data for which the solar wind bulk 
speed is larger than 650 km/s. The extended data set analyzed here includes the 
events detected by three probes, i.e., Helios, Cluster II and Ulysses, 
on an interval covering a radial distance of $0.3 - 4$ AU and without any 
restriction to fast or slow wind events.

\begin{table}[h]
\caption{Mean values for the solar wind speed (displayed in Fig.~\ref{f2}), and the 
parallel and perpendicular temperatures (displayed in Fig.~\ref{f3}) as determined for each of 
the core and halo components (same units as in Table~\ref{tab1}). 
The intervals for the $\kappa$-index values (bins) adopted in the calculation are the following: 
$<$2.75; 2.75-3.25; 3.25-3.75; 3.75-4.25; 4.25-4.75; 4.75-5.25; 5.25-5.75; 5.75-6.25; 6.25-6.75; 6.75-7.25;
7.25-7.75; 7.75-8.25; $>$8.25}
\label{tab2}
\begin{tabular}{c c c c c c c c c c c c c c}\hline
$\kappa$ &  2.5 & 3. & 3.5 & 4. & 4.5 & 5. & 5.5 & 6. & 6.5 & 7. & 7.5 & 8. & 8.5\\
\hline $u$ & 479 & 434 & 442 & 460 & 477 & 482 & 468 & 456 & 449 & 438 & 419 & 395 & 373 \\
\hline $T_{c,\parallel}$ & 1.07 & 1.18  & 1.26 & 1.35 & 1.35 & 1.33 & 1.32 & 1.32 & 1.37 & 1.47 & 1.65 & 1.81 & 1.87 \\
\hline $T_{c,\perp}$ & 0.85 & 1.02 & 1.08 & 1.13 & 1.12 & 1.12 & 1.13 & 1.15 & 1.21 & 1.32 & 1.52 & 1.72 & 1.80 \\
%\hline $A_c$ & 0.79 & 0.86 & 0.86 & 0.84 & 0.84 & 0.84 & 0.86 & 0.87 & 0.89 & 0.90 & 0.92 & 0.95 & 0.96 \\
\hline $T_{h,\parallel}$& 13.7 & 11.7 & 10.1 & 9.32 & 8.49 & 7.64 & 7.04 & 6.57 & 6.24 & 6.10 & 6.06 & 6.02 & 6.04\\
\hline $T_{h,\perp}$& 13.1 & 11.2 & 9.58 & 8.54 & 7.69 & 7.08 & 6.63 & 6.24 & 6.04 & 6.00 & 5.99 & 5.99 & 6.06\\
%\hline $A_h$ & 0.95 & 0.96 & 0.94 & 0.92 & 0.90 & 0.93 & 0.94 & 0.95 & 0.97 & 0.98 & 0.99 & 1.00 & 1.00 \\
\hline
\end{tabular}
\end{table}

Fig.~\ref{f1} displays the radial profiles 
of densities, temperatures and the power-index $\kappa$, which are
computed distinctively for the core (cross) and halo (star) components. This 
enables direct comparison with the results in \inlinecite{ma05}, e.g., 
Fig.~4 therein. 
We have found that the electron densities of the core and halo populations 
exhibit the same systematic decrease with heliocentric distance, but, contrary to the
results in \inlinecite{ma05} the temperatures of these two components 
show an opposite evolution with radial distance. While the core
temperature decreases and show a tendency of stabilization after 2~AU 
(when probably the expansion is considerably slowed down and so is its 
influence on the temperature), the halo temperature increases with an apparent
saturation of about 3~AU (radial bin 7). After 3~AU the halo temperature starts 
to decrease but the data from larger distances after 4~AU are not 
available to confirm this tendency. The decreasing gradient found for the 
core temperature is similar to those reported for the electron temperature in general
in the ecliptic and at higher latitudes, see \inlinecite{is98}, \inlinecite{ma05} or 
\inlinecite{ma00} and Table~1 therein.

For the halo temperature we have found a slow trend of increasing with heliocentric
distance which was not reported before, and is opposite to the 
gradients of decreasing obtained by the previous studies mentioned above. 
However, in all these studies, excepting \inlinecite{ma05}, the halo component was 
neglected or described only partially by a standard Maxwellian. In \inlinecite{ma05} 
the halo is accurately described by a bi-Kappa distribution function, and the halo
temperature is found to have a nonmonotonic evolution with heliocentric distance, 
which is presented in their Fig.~4b. In that figure, a similar slow trend of increasing 
of the halo temperature with heliocentric distance can be observed for a shorter range of 
the ecliptic data: If we exclude the 5th (last) bin (1.35-1.5~AU) which corresponds to 
the Ulysses data from very high latitudes, the increase of the halo temperature becomes 
evident between the radial bins 2 (0.475-0.53~AU) and 4 (1~AU). Otherwise, if we consider
the 5th bin, the resulting overall evolution of the halo temperature will indicate
a decreasing with heliocentric distance, and this was the conclusion reached by 
\inlinecite{ma05}. However, the 5th bin represents in their case the high latitudes data 
that may be less relevant for the ecliptic.

\begin{figure}[h]
%\centerline{
    \includegraphics[width=60mm]{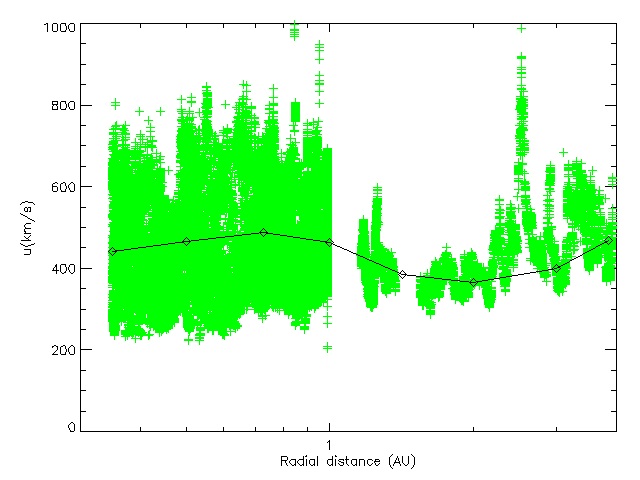}
    \includegraphics[width=60mm]{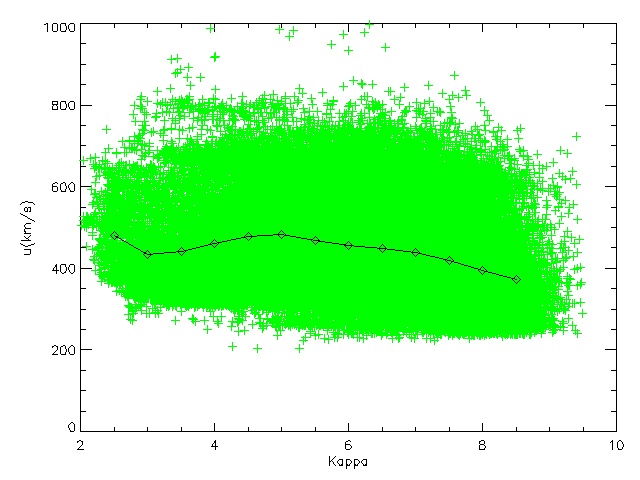}
\caption{The solar wind flowing speed as a radial profile (left) and a function of $\kappa$ (right).}
\label{f2}
\end{figure}

Corresponding to the increase of the halo temperature, the power-index $\kappa$ is
diminished with heliocentric distance from an average value between 7 and
8 at 0.3~AU to about $\kappa \simeq 3$ at 3~AU. The power-index $\kappa$ from the data 
set in \inlinecite{ma05} shows the same evolution, with mention that in their study 
the radial bin~5 provides information from very high latitudes.
Both evolutions given by the decrease of $\kappa-$index and the increase of
the halo temperature are in agreement with the enhance of halo populations with 
radial distance concluded by \inlinecite{ma05}.

In addition, in Fig.~\ref{f2} we display both the radial profile (left panel) and the 
$\kappa$-dependency (right panel) of the solar wind flowing speed, which show a nonmonotonic
variation. After the acceleration trend up to 0.7~AU, the solar wind speed decreases until it
shows pronounced peaks at large distances $> 2$~AU, most probably caused by some energetic events 
like coronal mass ejections. To keep this observational evidence, in Fig.~\ref{f2} the average values are superimposed
over the scatter plot data. The right panel in the same figure confirms a tendency of increasing
of the power-index $\kappa$ in the slow solar winds, as observed by \cite{Pier01}. 
In our case $\kappa$ increases to values exceeding 9, when the Kappa distribution function 
approaches the Maxwellian limit, see the two-Maxwellian models used by \inlinecite{ma00}.
%

\begin{figure}[h]
%\centerline{
    \includegraphics[width=60mm]{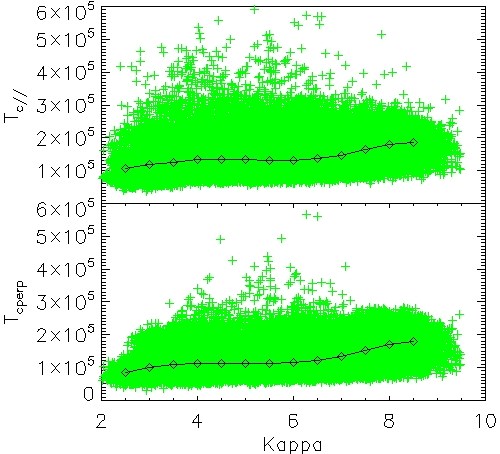}
    \includegraphics[width=60mm]{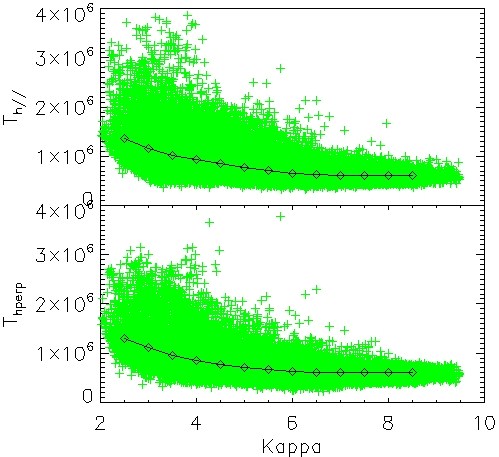}
\caption{Temperature components parallel (top) and perpendicular (bottom) to the
magnetic field vs. kappa index for the core (left) and halo (right) populations.} \label{f3}
\end{figure}

%In the next, a particular attention is given to the temperature and the anisotropy of the electron populations. 
Fig.~\ref{f3} displays the temperature components, parallel (top panels) and perpendicular 
(bottom panels) to the magnetic field direction, for both the core (left panels) and halo populations (right panels). The core show a modest increase of both the temperature components with the $\kappa$-index, while the halo
seems to compensate, exhibiting a very clear decrease of the parallel and perpendicular temperatures with 
increasing $\kappa$. These evolutions of the halo temperatures appear to be in perfect agreement with
the recent theoretical studies \cite{la15b}, which indicate an increase of the Kappa temperature with a decrease
of $\kappa$, i.e., an enhance of suprathermal populations, rather than a $\kappa$-independent temperature
\cite{li13}.  
An increase of $T_h$ with decreasing $\kappa$ is also in accord with 
the radial profiles of these two quantities provided by our analysis above. Moreover, such a 
$\kappa$-dependence of the core and halo temperatures clarifies an apparent contradiction
in \inlinecite{ma05}, where the halo temperature is found to decrease with radial distance at the same time
with an increase of the power-index $\kappa$. As discussed above, the radial decrease of $T_h$ 
found by \inlinecite{ma05} is given by the high-latitude data, most probably with a lower relevance
for the radial evolution in the ecliptic. These comparisons indicate a decrease of $T_h$ at 
higher latitudes, but a rigorous analysis needs an extended mapping of the Ulysses data.

\begin{figure}[h]
\centering
    \includegraphics[width=62mm]{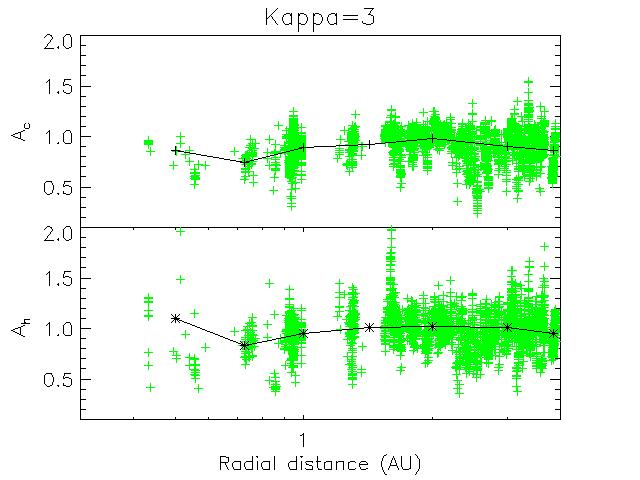} \includegraphics[width=55mm]{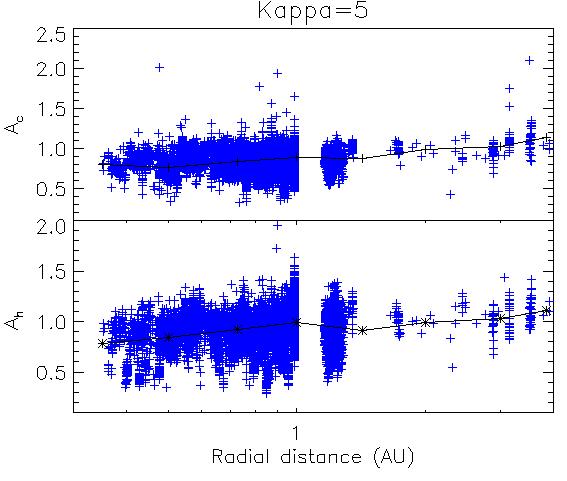}\\
    \includegraphics[width=62mm]{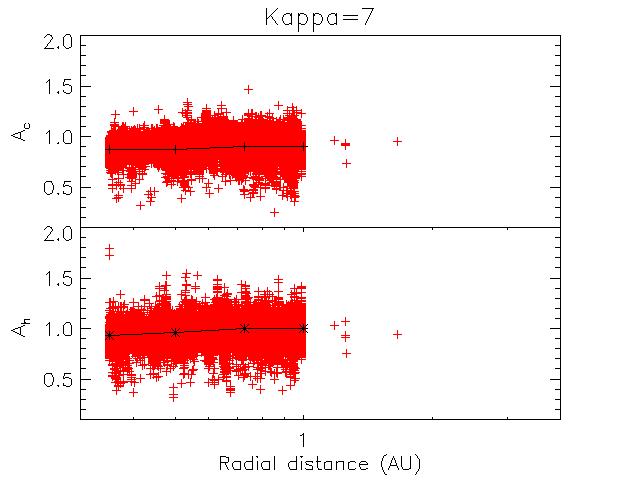}
\caption{Radial profiles for the temperature anisotropy for three classes of events associated to three
distinct values of the power index: $\kappa \simeq 3 \in [2.7, 3.3]$ (top-left panel), $\kappa \simeq 5
\in [4.5, 5.5]$ (top-right panel) and $\kappa \simeq 7 \in [6.3,
7.7]$ (bottom panel).}
\label{f4}
\end{figure}

From a radial profile of the power-index $\kappa$ in Fig.~\ref{f1}, 
bottom-right panel, we may identify three groups 
of events associated to three distinct (median) values of the
power-index as follows: $\kappa = 7$ typical for low heliocentric
distances $R < 0.7$~AU, $\kappa = 5$ specific for intermediary
distances 0.7~AU$ < R < 1.5$~AU, and $\kappa = 3$ representative for
large radial distances $R > 1.5$~AU. From the total number of
events we select for further analysis only those associated with these 
three characteristic values of the power-index $\kappa = 3$, 5, and 7. 
Thus, Fig.~\ref{f4} presents a radial evolution of the temperature
anisotropy for each of these three classes of events, building their radial profiles and
making a comparison between the temperature anisotropy of the core and halo populations. 
Although these populations show deviations from isotropy in both directions, namely, $A = T_\perp /T_\parallel >1$ 
due to an excess of perpendicular temperature, or $A = T_\perp /T_\parallel < 1$ by a
surplus of parallel temperature, dominant seems to be the number of states with $A < 1$.
The mean values of the temperature anisotropy (superimposed over the 
scatter data points in Fig.~\ref{f4}) do not deviate much from the conditions
of isotropy $A_c = 1$ and $A_h = 1$, and have therefore a reduced relevance. Relevant in this case is the spread of 
data points, which is larger, sometimes markedly larger for the halo than for the core populations.
Deviations from isotropy decrease with increasing $\kappa$, being thus larger for a lower $\kappa =3$
while the higher $\kappa = 7$ populations are narrower and more compact.

\begin{figure}[h]
%\centerline{
    \includegraphics[width=60mm]{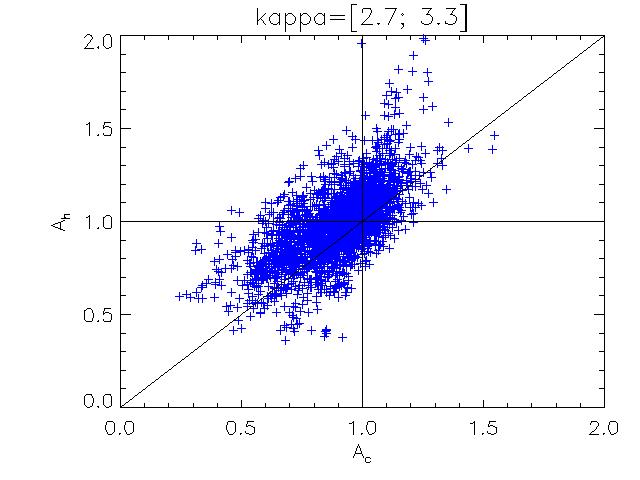}
    \includegraphics[width=60mm]{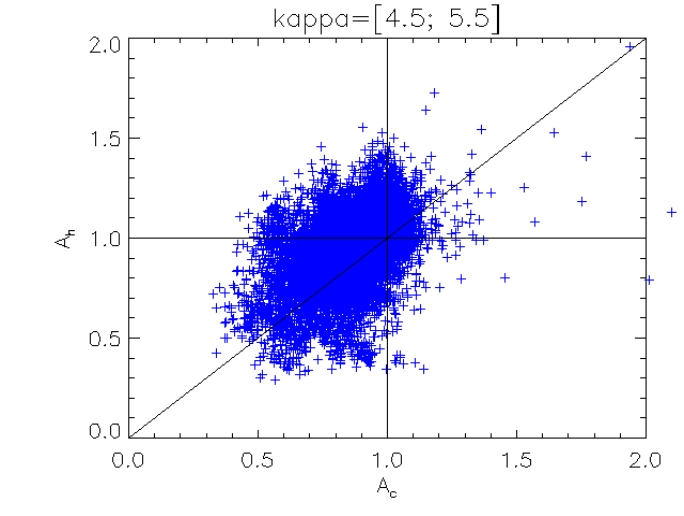}\\
    \centering \includegraphics[width=60mm]{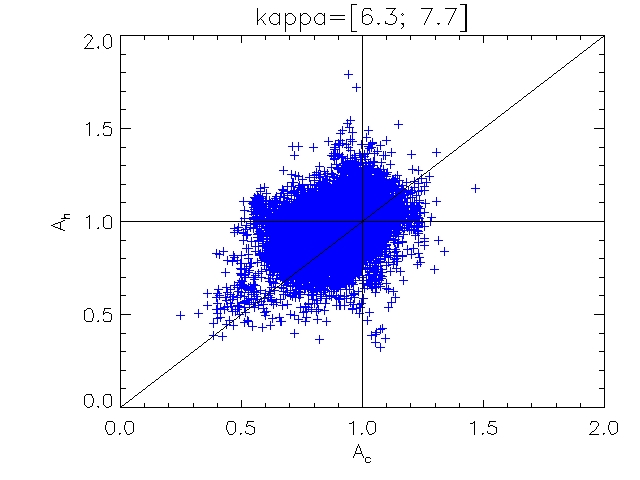}
\caption{The electron temperature anisotropies $A_h$ vs. $A_c$ for
the same classes of events chosen in Fig.~\ref{f4}.} \label{f5}
\end{figure}

In order to correlate the anisotropies of the core and halo populations and have a better 
image of their spread, in Fig.~\ref{f5} we plot the halo anisotropy vs. the core anisotropy
for each of these sets of events with the power-index in the range $\kappa = 3 \pm 10\% \in [2.7, 3.3]$ in
panel a, the events with $\kappa = 5 \pm 10\% \in [4.5, 5.5]$ in panel b, and the events with $\kappa = 7 \pm 10\%
\in [6.3, 7.7]$ in panel c. The data points are not scattered uniformly in the four quadrants delimited
by the isotropic temperatures for the core $A_c = 1$ and halo $A_h = 1$, and show a dominant presence
in the left quadrants defined by $A_c < 1$, and especially in that of the left-bottom corner where both 
the core and halo anisotropies, i.e., $A_c < 1$ and $A_h < 1$, are favorable to an excess of parallel 
temperature. On the other hand, the data points in all these three panels show a prevailing 
disposition of aligning to a straight-line (linear regression) which connects the origin of axes at 
$A_c = 0$ and $A_h = 0$, with the point defined by the isotropic states for both components $A_c = 1$ and $A_h = 1$. The spread of 
data points around this line becomes more balanced with increasing $\kappa$-index. This linear regression 
suggests a linear correlation between the core and halo anisotropies of a form $A_h = a A_c + b$, with 
$a \simeq 1$ and $b \simeq 0$, by which these two components manifest a tendency
to deviate from isotropy in the same direction, either the predominant states with both $A_c < 1$ 
and $A_h <1$, or the opposite states with both $A_c > 1$ and $A_h >1$. Apparently, this correlation 
of the data points becomes less pronounced with the increase of kappa index, e.g., $\kappa =$5, 7, 
when  their tendency of aligning is reduced but still distinguishable. However, this can be only a 
consequence of the increase of data points for higher values of kappa index, especially for $\kappa =$5. 
With the increase of $\kappa$, the data points also show a concentration towards the stability states 
of isotropy and thermal equilibrium for both components, i.e., $A_c = 1$ and $A_h = 1$. 
This may be a natural predisposition as the increase of $\kappa$ indicates a thermalization of the 
halo component most probably under the effect of collisions and fluctuations, which may reduce both the 
suprathermal populations and the temperature anisotropy. Otherwise, noticeable is also the presence of
anticorrelated states when the core and halo anisotropies are opposite, either $A_c> 1$ and $A_h <1$, 
or $A_c <1$ and $A_h >1$. The existence of these anticorrelated states may have many implications, 
suggesting, for instance, different origins of the core and halo components \cite{sc92,Pier99}, but also a dynamic 
interplay of these populations mediated, most probably, by the kinetic instabilities \cite{la14, sh16}.

%%%%%%%%%%%%%%%%%%%%%%%%%%%%%%%%%%%%%%%%%%%%%%%%%%%%%%%
\section{Discussions and conclusions}
%%%%%%%%%%%%%%%%%%%%%%%%%%%%%%%%%%%%%%%%%%%%%%%%%%%%%%

Estimating the anisotropy of the electron temperature in the solar wind has a 
particular importance for understanding the implications as well as the physical 
mechanisms which are responsible for these deviations from thermal equilibrium, e.g., the solar wind 
expansion, or the heating of plasma particles by the fluctuations. 
Large deviations from isotropy are local sources of instabilities that may enhance the fluctuations, 
and act in turn, pitch-angle scattering the particles and limiting the increase of
temperature anisotropy. On the other hand, the solar wind is a hot and dilute plasma, where 
particle-particle collisions are rare and therefore expected to maintain close to the 
thermal equilibrium only the low-energy core populations. This is confirmed by the observations,
which also show that large deviations from temperature isotropy appear to be better constrained 
by the resulting instabilities \cite{st08}. 
More intriguing are the suprathermal populations, which are less dense than the core, 
but hotter, such that their kinetic energy density is comparable to that of the 
core, i.e., $\beta_h \sim \beta_c$, see in \inlinecite{ma00} and \inlinecite{la15}. 
In these preliminary estimations both the core and 
halo populations are interpreted with idealized Maxwellian models, but the results 
strongly point out that these two components may have similar kinetic effects,  
and neither of them can be ignored in the favor of the other. Decoding  
the interplay of the thermal core and the suprathermal halo seems to be vital 
for a correct understanding of the kinetic mechanisms in the solar wind.

In this paper we have analyzed both these two components by contrast, focusing on their 
temperatures and temperature anisotropies. The data set analyzed here includes more than 120$\,$000
events detected in the ecliptic from 0.3 to over 4~AU by three spacecraft missions. 
According to the previous analysis \cite{ma05} both the core and halo temperatures are expected to 
decrease with heliocentric distance, but from our data set the halo temperature is found to be 
enhanced with increasing the distance from the Sun. This apparent contradiction can be 
however explained by the difference between the data sets considered in these investigations.
Thus, our data set restricts to the events in the ecliptic, while the previous analysis 
combines data from the ecliptic with high latitude data. Without these data collected 
by Ulysses from over the solar poles, the halo temperature in the ecliptic also show a 
radial increasing similar with the one found in the present paper.
On the other hand, the power-index $\kappa$ is found to be diminished with heliocentric distance, 
suggesting that the core temperature $T_c$ and the halo temperature $T_h$ must show opposite 
variations with the $\kappa$-index, i.e., a decrease of $\kappa$ should determine an increase of $T_h$,
while $T_c$ is diminished in the same measure. These variations are indeed confirmed by the 
observations in Fig.~\ref{f3}, and they come to provide an important observational support 
for the recent studies \cite{la15b} which suggest that Kappa modelling of the plasma particles  
and their kinetic effects in the solar wind must consider a $\kappa$-dependent temperature,
increasing with decreasing the $\kappa$-index.

Radial profiles of the temperature anisotropies have been built in Fig.~\ref{f4} for three 
groups of events associated to distinct values of the power-index $\kappa = 3, 5, 7$, where the lowest 
value is found representative for large distances from the Sun $R > 1.5$~AU, and the 
highest value of $\kappa$ for low distances $R<0.7$~AU. Relevant is the spread of data points
indicating deviations from isotropy, which in many situations are markedly larger for the halo than
for the core. These deviations from isotropy decrease with increasing $\kappa$, such that the higher
$\kappa = 7$ populations are more compact along the isotropy condition ($A=1$) than those modelled by 
lower values of $\kappa$. We have also looked for a direct interdependence between the core and 
halo anisotropies, building panels in Fig.~\ref{f5}, again for the same groups of events associated 
with three distinct values of the power-index $\kappa$. The core and halo components manifest a clear 
tendency to deviate from isotropy in the same direction, confirming the existence of 
mechanisms of energization or relaxation with the effects on both components, e.g., the solar wind 
expansion, or the particle heating by the fluctuations. Moreover, the existence of plasma states with 
anti-correlated anisotropies of the core and halo populations suggests a dynamic interplay of these 
components, mediated, most probably, by the anisotropy-driven instabilities. Indeed, for the core population 
the large deviations from isotropy appear to be constrained by these instabilities \inlinecite{st08}.
For the halo populations such a study does not exist and it will make the object of our next 
investigations.

%We have not taken into discussion the field-aligned strahl, which usually is 
%less dense than the core and halo populations and becomes significant during the
%energetic events (fast winds and CMEs). 
%The results in the present paper are representative for an extended
%range of heliocentric distances, i.e., 0.3 - 4~AU, in the ecliptic.

\begin{acks}

The research leading to these results has received funding from the
Scientific Federal Policy in the framework of the program
Interuniversity Attraction Pole for the project P7/08 CHARM.
The authors acknowledge support from the Katholieke Universiteit Leuven,
the Ruhr-University Bochum, and Alexander von Humboldt Foundation. 
These results were obtained in the framework of the projects
GOA/2015-014 (KU Leuven), G0A2316N (FWO-Vlaanderen), and C 90347
(ESA Prodex 9). The research leading to these results has also
received funding from the European Commission's Seventh Framework 
Programme FP7-PEOPLE- 2010-IRSES-269299 project-SOLSPANET
(www.solspanet.eu). The authors further acknowledge the grant 15-17490S 
of the Czech Science Foundation.

\end{acks}
%\href{}{

%%%%%%%%%%%%%%%%%%%%%%%%%%%%%%%%%%%%%%%%%%%%%%%%%%%%%%%%%%%%%%%%%%%%%%%

%%%%%%%%%%%%%%%%%%%%%%%%%%%%%%%%%%%%%%%%%%%%%%%%%%%%%%%%%%%%%%%%%%%%%%%

\end{article}
\end{document}